\documentclass{knawproc}
\usepackage{epsfig}

\begin{document}

\begin{opening}

\title{Faint radio samples: the key to understanding radio galaxies}

\author{Stephen A. Eales}
\addresses{%
  Department of Physics and Astronomy, University of Cardiff, P.O. Box 913,
Cardiff CF2 3YB, UK\\
}

\end{opening}


\begin{abstract}

The large number of differences between high- and low-redshift radio galaxies
have almost all been discovered by looking at the bright 3C sample of radio
sources. This has the disadvantage that the strong correlation between radio
luminosity and redshift within a single sample makes it impossible to
be determine whether these differences are the result of cosmic evolution
or whether they are simply the result of source properties depending on radio
luminosity.
The solution to this problem
is to compare the properties of sources in faint samples with those of
the 3C sources. I and collaborators have recently removed the
degeneracy between redshift and radio luminosity by comparing the
3C sample to the recently completed 6C and 7C samples.
In this paper I concentrate on what our study has revealed about the
host galaxies of radio sources. At low redshift, radio galaxies are giant
ellipticals with absolute magnitude being independent of radio luminosity
over a range of $\rm 10^4$ in radio luminosity. At $\rm z \sim 1$, the 
radio-luminous 3C radio galaxies are still giant ellipticals, but the 6C 
galaxies, only a factor of
six lower in radio luminosity, are fainter by about 1 mag in the near-IR and
have much more compact near-IR structures. At $\rm z \sim 1$, radio galaxies
follow a line in a diagram of optical luminosity verses de Vaucouleurs scale
length parallel to the projection of the fundamental plane 
for nearby ellipticals in this diagram. I discuss the significance of 
these results for
our understanding of radio galaxies and their evolution.

\end{abstract}


\section{Introduction}

There are sixty participants in this conference. As about half of my 
own circle
of radio galaxy enthusiasts is here, I estimate that, in the whole world, 
there
are about one hundred people interested in high-redshift radio galaxies. 
I suspect that in the nineteen eighties this number would have been higher, 
because, for a few years, radio galaxies were very fashinable, being then 
the only galaxies one could easily observe at high redshift and thus offering 
the only prospect of
investigating the early evolution of galaxies. Well, fashion has moved on. 
The advent of faint galaxy redshift surveys, such as the Canada-France 
Redshift Survey, which have found large numbers of normal galaxies out to 
cosmologically-useful redshifts, the Hubble Deep Field, the successful 
implementation of the Lyman-break technique---all these have pushed radio 
galaxies off the catwalk. Radio galaxies are now recognised for what they 
really are: awkward objects that are neither entirely galaxy nor entirely 
quasar, objects whose properties at high redshift have a strictly limited 
applicability to the evolution of galaxies in general (except perhaps in 
the case of those objects with the lowest radio luminosities---Dunlop, this 
meeting). I actually find this new unfashionableness rather appealing. 
Since radio galaxies are interesting objects in their own right, it is quite
pleasant not to have to strain to make tenuous connections to general issues 
of galaxy evolution. Paradoxically, however, it is just at the moment that 
radio galaxies seem deeply out of fashion that we are beginning to make 
significant progress in understanding radio galaxies as active galaxies, and 
I even have the suspicion that this may yet have relevance to broader 
aspects of galactic evolution.
 
Much of this recent progress has been made by studying faint samples 
of radio sources. As everyone at this conference knows, one of the 
biggest individual contributions to our field was made by Hyron 
Spinrad in his spectroscopic observations of the 3C sample in the seventies 
and eighties (the ''Spinrad era'', in George
Miley's words). Until recently the 3C sample$^{1,2}$ was 
the only sample of radio 
sources with almost complete redshift information, and most of the very 
significant differences between high- and low-redshift radio galaxies 
(HZRG's and LZRG's, for short) were 
discovered from observations of the 3C sample. There are six differences 
I can think of:
(1) HZRG's have higher near-IR 
luminosities$^3$; (2) HZRG's have bluer optical-near-IR colours$^3$; 
(3) LZRG's and HZRG's have radically different structures, the former being 
relatively normal giant ellipticals, the latter having typically a 'bead of 
pearls ' structure$^{4,5}$ (HST observations have shown that
this phrase does not do justice to the very bizarre structures of HZRG's, 
which are impossible to sum up verbally, but which anyway look nothing 
like that of a giant elliptical$^6$); (4) The optical continuum structures of 
the HZRG's , but not those of the LZRG's, are aligned with their radio 
structures$^{4,5}$; (5) The emission-line luminosities of the HZRG's are 
higher than those of the LZRG's$^{7,8}$, and the line-emitting 
gas is frequently 
found in very extensive ($\rm D \sim\ 100\ kpc$) nebulae$^9$; (6) The 
physical sizes of the radio sources associated with the HZRG's
are smaller, on average, than those associated with the LZRG's$^{10,11}$.

Thus there are many differences between radio galaxies at high and 
low redshift, but our reliance on a single flux-limited sample (within which 
radio luminosity and redshift are tightly correlated---Fig. 1) has meant that 
it has been impossible to distinguish whether these differences are genuinely 
due to the effect of redshift (cosmic evolution) or are caused by the 
different radio luminosities of
3C radio galaxies at high and low redshift.
As I will show, the way to determine whether we are seeing cosmic evolution
or a luminosity effect is to obtain spectroscopic redshifts for faint
samples, and for over ten years Steve Rawlings and I and a large number
of collaborators (see acknowledgements) have been trying to obtain
redshifts for faint samples selected from the Cambridge 6C and 7C radio
surveys. One of the several advantages of these surveys is that they
were carried out at 151 MHz, very close to the frequency of the 3C survey
(178 MHz), thus reducing the well-known problem that the mix of 
radio morphologies in a radio sample,
the distribution of radio spectral indices, and the proportions of the
sources that are quasars and galaxies is a strong function of the selection
frequency. There are several samples. The 6C `2-Jy' sample has 
flux limits of approximately 2 and 4 Jy, the lower flux
limit being about six times fainter than that of the 3C sample, and consists of 
64 sources. Of these sources, only
two (one of which is very close to a star) do not yet have redshifts.
The 6C sample overlaps in area with the sample of Allington-Smith$^{12}$
which was selected from the 408-MHz B2 survey. The combined 6C/B2 sample
contains 80 sources. All the infrared imaging data for these samples
has now been published$^{13}$ and forthcoming papers will present the
redshifts and technical details about the samples. There are various
7C samples (Lacy, this meeting), which are all about four times fainter
than the 6C sample. These samples have been mainly studied by the Oxford
group, and there are now redshifts for $\simeq$ 90\% of the sources.
The 7C samples are even a shorter distance along the road to publication
than the 6C sample. These faint samples are the only ones of which I am aware
which have such complete redshift information.

\begin{figure}[h]
\vspace{5cm}
\caption{Radio luminosity at 151 MHz in $\rm W\ Hz^{-1}\ sr^{-1}$
verses $\rm (1+z)$ for the 3C sample$^1$ (open circles) and the 6C sample
(filled circles). A Hubble
constant of 50 $\rm km\ s^{-1}\ Mpc^{-1}$ and a density constant of 1 have
been assumed.
}
\end{figure}

Figure 1 shows the 3C and 6C samples plotted on the 
radio luminosity-redshift plane. As for 3C, within the 6C sample 
there is a tight correlation
between redshift and radio luminosity.
However, the combination of samples allows one to disentangle the effects
of redshift and radio luminosity. At any redshift, the combination of the
samples provides a sufficient
range of radio luminosity for one to look for correlations between a
third property and radio luminosity.
Similarly, there is a large range of redshift at 
any radio luminosity for one to look for the effects of cosmic evolution
without the worry that any effect could be caused by the radio luminosities
of the high-redshift sources being different from those at low redshift.
The 7C sources fall below the 3C and 6C sources in this diagram, extending
the range of radio luminosity at constant redshift and the range of
redshift at constant radio luminosity.

\section{What we have learned from the faint samples}

Our study of the faint low-frequency samples has allowed us to answer
three basic questions about radio galaxies.

\subsection{Why is it hard to measure redshifts for faint radio sources?}

Studies of the 3C sample$^{7,8}$ 
have shown that emission-line
luminosity and radio luminosity are strongly correlated, but
without considering fainter samples
it is impossible to determine whether this is the true correlation or
whether the true correlation is between emission-line luminosity and redshift.
Although a full quantitative analysis of our 6C spectra is still in progress,
it is already obvious that 6C galaxies have weaker lines than 3C galaxies
at a similar redshift. This shows that the true correlation is between
emission-line luminosity and radio luminosity.
One plausible (although not unique) explanation of this correlation is 
that both the emission-line luminosity
and the radio luminosity are measuring the
`power of the central engine', the emission-line luminosity because
it is from gas
which is being photoionized 
by radiation from the active nucleus and the radio
luminosity because it is a strong function of the kinetic power
of the beam being produced by the active nucleus$^{14}$.
From the point-of-view of an observer, the message of the correlation 
between radio and emission-line luminosity is that there is a limit to how faint
one can go in radio flux and still obtain complete redshift information.
The 7C sample is about twenty times fainter than 3C, and we believe that
this is as faint as is practical to go with 4-m telescopes. Any fainter,
as Jim Dunlop will show in his talk, and you are in the regime of the
8- and 10-m telescopes.

\subsection{Do the sizes of radio sources change with redshift?}

One of the earliest discoveries about high-redshift radio sources
was the discovery that the radio sizes of high-redshift quasars
are smaller than those of quasars at low redshift$^{10,11}$ (I would
be interested to know if there are any earlier references on this
subject than 1970). 
Of course, because
of the correlation between radio luminosity and redshift within a bright
sample, it is possible that this discovery actually means that
radio sources with high radio luminosities have smaller radio sizes
than those with low radio luminosities. Since those early days there
has been a veritable industry trying to disentangle the correlations
and to determine the strength of the evolution (if it is evolution).
In principle, by comparing the sizes of sources from the 6C and 7C
samples with the sizes of sources of similar luminosities in the
3C sample, it should be possible to get an unambiguous measurement
of the strength of any evolution, and by comparing 6C/7C sources
with 3C sources at a similar redshift it should be possible to
look for correlations of size with radio luminosity. One of the
biggest problems in this kind of analysis is that the sizes of sources
depend on the selection frequency of the sample in which they
were found, because high-frequency samples tend to contain more
compact sources (both steep spectrum and flat spectrum). The closeness
of the selection frequencies of 3C, 6C and 7C means that, for avoiding
this problem, this combination of samples is almost but not quite ideal.
That the combination is not quite ideal is because of a fairly subtle
point. Suppose one has selected a sample at a particular frequency, the
frequency at which the radiation was emitted by one of the sources
in the sample will have been higher than the selection frequency 
by a factor $(1 + z)$. If one is comparing sources of a similar
radio luminosity in a bright and a faint sample, the $1+z$ factor will
be larger for the faint sample than the bright sample, meaning that
the {\it effective} selection frequency of the faint sample will
be higher than that of the bright sample even if the actual selection
frequencies were the same, raising the possibility again that spurious
cosmological evolution could be seen because of the proportion of
compact sources increasing with selection frequency. Thus it would be
preferable if the selection frequency of the 6C/7C samples was even
lower than 151 MHz. Nevertheless, for a comparison with 3C, the low
selection frequency and the large percentage of redshifts means
that these faint samples are the best available.

\begin{figure}[h]
\vspace{9cm}
\caption{Radio luminosity at 151 MHz in $\rm W\ Hz^{-1}\ sr^{-1}$
verses verses source size in kpc for the 3C sample (a) and for the
6C sample (b). On both graphs the filled circles represent FR2 sources
and the open circles sources with other morphologies.
The horizontal lines indicate the approximate luminosities that
sources with redshifts of 0.5, 1, and 2 would have in the two samples.
}
\end{figure}

A couple of years ago, Mark Neeser and I did a linear-size 
analysis$^{15}$ using 
the 6C
and 3C data and found rather weaker 
cosmological evolution ($D_{med} \propto (1+z)^{-1.5}$) than had
previously been claimed and no evidence
for a correlation between radio size and radio luminosity. 
This evolution can be seen visually in the P-D diagrams of the two
samples (Fig. 2) 
by looking at the `clump' of sources in the 6C diagram with sizes
of about 100 kpc and luminosities 
between $\rm 10^{27}\ and\ 10^{28}\ W\ Hz^{-1}\ 
sr^{-1}$. Most of the sources in this clump actually have sizes less
than 100 kpc, whereas in the same luminosity range in the 3C diagram
there is much larger fraction of sources with sizes greater than this.
We found the apparent clump quite intruiging
because, as in the case of the H-R diagram, the distribution
of objects in the P-D diagram should reflect the relative times
that sources spend at different stages of their evolution$^{16,17}$. 
Furthermore the
physical sizes of the clump sources are very similar to the typical sizes
of the gaseous nebulae found around HZRG's, and we constructed a simple
model in which the evolution of a high-redshift source is strongly affected
by the presence of one of these nebulae.

At this time we thought we had had the last word
on the subject, both because of the
quality of our data and because we had spotted a hitherto
unrecognised selection effect which we thought had caused other groups
to detect too strong evolution. It was therefore quite stimulating to
hear Pat McCarthy at this meeting say that, on the basis of the MRC 1-Jy sample,
he and his collaborators see much stronger linear-size evolution
than we found, a 
correlation
between radio size and radio luminosity, and no clump. How do we resolve this
disagreement? I am suspicious that the differences are caused by the 
selection-frequency effect. As discussed above, the effective difference
between the selection frequencies of the MRC and 3C samples is greater
than the difference between the actual frequencies (408 MHz and 178 MHz),
and this could produce a larger fraction of compact sources in the
MRC sample than we find in the 6C sample. Fortunately, the quality of
the data for all faint samples is now such that we can go beyond mutual
suspicion. By comparing the P-D diagrams of the 6C, 7C and MRC samples,
and in particular by plotting the P-D diagrams for the various morphological
classes within these samples, we should be able to determine immediately
why we obtain different results. Overall, I think there is the prospect
of a rapid advance in this area,
since for the first time we can plot 
P-D diagrams for samples of different flux densities
and selection frequencies
which 
are not missing high-redshift sources and sources of large angular size---two
problems that bedeviled early investigations in this area. Our theoretical
understanding of the P-D diagram is also advancing at a satisfactory
result. We are beginning to understand the evolutionary connections
between different morphological classes$^{18}$, and there are finally
theoretical models that predict, in a natural way, some of the gross features
of the P-D diagram$^{19}$.

\subsection{What kind of galaxies are high-redshift radio galaxies?}

One of the few unassailable facts 
in our field is that at low redshift radio galaxies
are giant ellipticals with a small spread of absolute magnitude. 
It is worth reminding ourselves that, beyond some hand-waving
theorizing, we still do not understand why this should be. When one
thinks about it, it is quite surprising that, over a range of $\rm 10^4$ in
radio luminosity, there is no clear relation between radio and optical
luminosity$^{20}$. Although optical images of HZRG's look nothing like
giant ellipticals, until recently it had been possible to hope that
HZRGS's are still essentially giant ellipticals, with the optical 
emission being due to
`fireworks' occurring in the rest-frame
ultraviolet, either nonstellar emission from the active nucleus
or emission from star-formation regions (producing a lot of
light but of relatively low mass compared to the mass of the galaxy as
a whole)$^{3,21}$. The true test of this idea is to make observations in the 
near-infrared, since these will be sensitive to the old stellar 
population and will be relatively unaffected by nonstellar light
or light from young high-mass stars. These observations have recently
shown that this basic fact may be true at $\rm z = 0$, but it is
not true at $\rm z \sim 1$.

A couple of years ago, when we started measuring K-magnitudes for
6C galaxies we were surprised to discover that 
at $\rm z \sim 1$ 6C galaxies are systematically fainter by
about 0.6 mags than 3C galaxies at similar redshifts$^{13}$. This
is quite a remarkable result, and very different from the situation
at low redshift, since it shows there is a 
correlation between radio and near-IR luminosity over a range
of only $\simeq$6 in radio luminosity. Our first thought was that we were
seeing the effect of nonstellar emission. There is plenty of evidence
from polarization studies$^{22}$ of scattered nonstellar light from HZRG's
in
the optical waveband; if nonstellar light
is also making a significant contribution in the K-band (either
scattered light or emission directly from the active nucleus),
then one would expect 3C galaxies to be brighter in the near-IR than 6C
galaxies, because of their greater radio luminosities and thus presumably
more powerful active nuclei.
There is some evidence for nonstellar K-band emission from high-redshift
3C galaxies. In the K-band the narrow-line radio galaxy 3C 22 has the
properties of a quasar: a bright unresolved continuum source 
and broad lines$^{23}$. There is evidence in a few cases that the 
K-band light is polarized$^{24}$, suggesting a scattered component to
the K-band light. But the K-band morphology of 3C 22 is almost 
unique---virtually all 3C galaxies are extended in the K-band---and
the fraction of the K-band light that is estimated, from the polarization
measurements, to be scattered is insufficient to explain the large 
difference between the near-IR luminosities of the 3C and 6C galaxies.
The most conclusive arguments against this hypothesis, however, come from
recent high-resolution K-band imaging of 3C and 6C galaxies.

Best and collaborators$^{25,26}$ have recently shown that the
K-band structures of high-redshift 3C radio galaxies are well-fit
by the de Vaucouleurs profiles typical of giant ellipticals. Using the
REDEYE camera on the CFHT, we have carried out a similar imaging survey
of 6C galaxies at $\rm z \sim 1$ $^{27}$. We too find that the intensity 
profiles
can be adequately fitted by a de Vaucouleurs profile, but with the 6C
galaxies having, on average, a much smaller value of the de Vaucouleurs
radius than the 3C galaxies. Both sets of data are plotted in Fig. 3, which
shows that 6C galaxies at $\rm z \sim 1$ are both less luminous
in the K-band than 3C galaxies at a similar redshift and also have
more compact structures. The difference in luminosities is actually 
$\simeq$1 mag, which is greater than the difference we found before$^{13}$; 
the
discrepancy arising because in the former study we compared aperture
magnitudes, whereas the magnitudes plotted in the figure are total
magnitudes (see [27] for a discussion). The 3C and 6C HZRG's follow
a line in the diagram which
lies roughly parallel to the line followed by low-redshift ellipticals.

\begin{figure}[h]
\vspace{5cm}
\caption{Estimated half-light radius, which is here equal to the de Vaucouleurs
radius, verses optical luminosity for
high-redshift 6C galaxies (filled circles) and high-redshift 
3C galaxies (open circles). The dashed line shows the track in the 
diagram followed by nearby ellipticals. See [27] for a full discussion
of how this diagram was constructed.
}
\end{figure}

The study of Best et al. is conclusive evidence 
that in general the K-band light from 3C galaxies at $\rm z \sim 1$
is not dominated by nonstellar emission. Although our study of 6C galaxies
had similar
angular resolution ($\simeq$ 1\ arcsec) to the 3C study, the much more compact
structures of the 6C galaxies mean that our limits on the presence of nuclear
nonstellar sources are less stringent than the limits for the 3C galaxies.
However, if there were nuclear nonstellar sources in the 6C galaxies but
not in the 3C galaxies, this would mean that the difference between the
luminosities of the host galaxies would be even greater than is apparent
in Fig. 3. Moreover, it would be most surprising if the less radio-luminous
galaxies (and thus presumably the ones with the less powerful active nuclei)
had stronger nuclear nonstellar K-band sources; as well as being the opposite
of the explanation that we originally proposed for the difference
in the K-band luminosities. The compact structures
of the 6C galaxies mean that we can also not prove that these galaxies
are elliptical galaxies: the measured intensity profiles are consistent with
both de Vaucouleurs and exponential profiles. However, as in the case of the
nuclear nonstellar sources, it seems sensible to make the least
radical assumption. Therefore, we will assume, until shown otherwise, that 
the K-band emission from a 6C galaxy at $\rm z \sim 1$ follows a de Vaucouleurs
profile, and that there is negligible contribution from a nuclear source. 

Figure 3 immediately raises four questions: (1) What is the cause of the
correlation between radio luminosity and near-IR luminosity seen at $\rm z \sim 1$? (2) Why is such a correlation not seen at low redshift? (3) What will
the 6C galaxies at $\rm z \sim 1$ evolve into at $\rm z = 0$? (4) What will the
3C galaxies at $\rm z \sim 1$ evolve into at $\rm z = 0$? I will try and
answer the last two questions first, because I and my collaborators and
Philip Best and his collaborators have independently reached the same
conclusion. While not guaranteeing the solution is correct, it at least
ensures that nobody will disagree with me at this meeting. 

The dashed line in Figure 3 is a projection of the fundamental plane for
nearby elliptical galaxies.
As one would expect, low-redshift FR2 radio galaxies, being ellipticals,
lie approximately along this line$^{27}$, most of them having lower values of 
the de Vaucouleurs radius than the 3C HZRG's.
The 6C and 3C HZRG's must evolve
in such a way as to reach this line by the current epoch.  
There are two evolutionary mechanisms which will cause
galaxies to move across the diagram. Simple stellar evolution will cause
a galaxy to move horizontally across the diagram. If, as has been suggested,
most of the stars in a radio galaxy form at a very early time, the expected
amount
of passive stellar evolution between $\rm z = 1$ and $\rm z = 0$ is just
about enough to move the HZRG's onto the zero-redshift line$^{27}$. The
other type of evolution that can occur is merging. The effect of homologous
merging is
to make galaxies more diffuse and more luminous$^{28}$, moving them to
the top left in Fig. 1. This allows us to make a strong inference. Even if
there is no merging, 3C galaxies at $\rm z \sim 1$ {\it can not evolve
into low-redshift FR2's.} Their structures are already too extended (they
have too large de Vaucouleurs radii) at
$\rm z \sim 1$ compared with low-redshift FR2's. If there is any merging,
making the 3C galaxies even more extended, this conclusion is only strengthened.
So what do 3C HZRG's evolve into? A plausible step beyond this initial
conservative conclusion is to assert that 3C HZRG's evolve into first-ranked
cluster galaxies at the current epoch. There are two pieces of 
evidence for this. First, although there has been no systematic statistical
investigation of the environments of 3C galaxies at $\rm z \sim 1$, there
is evidence in many individual cases of surrounding clusters or of dense
surrounding gas$^{25}$. If a radio galaxy is in a cluster at $\rm z \sim 1$,
it will still be in a cluster at the current epoch. Second, the de Vaucouleurs
radii of 3C HZRG's are lower but not much lower than first-ranked cluster
galaxies, which means that it would not require much merging between 
$\rm z \sim 1$ and $\rm z = 0$ to give a 3C galaxy the intensity profile
of a first-ranked cluster galaxy. As the timescale for the evolution
of a radio source is only $\rm 10^8$ years$^{18}$, there is no reason
why a 3C HZRG need be a radio source at all at the current epoch and, as I have
argued, it can not turn into an FR2. It is possible, however, that the ultimate
descendant of a 3C HZRG could be an FR1, since these radio sources tend to be
found in clusters and are frequently associated with first-ranked 
cluster galaxies$^{29}$.

One can make a less definite answer to the question about the descendants
of 6C HZRG's. Since these have much lower de Vaucouleurs radii than the
3C HZRG's, they would have to undergo much more merging to turn into
first-ranked cluster galaxies at the current epoch. Because of the very
similar de Vaucouleurs radii, it seems most plausible that 6C HZRG's simply
turn into ellipticals like those that host FR2's at the current epoch.
If this is true, then as low-redshift FR2's tend to be found in quite
isolated environments$^{29}$, this suggests a way of testing the answers to 
both questions (3) annd (4). If these answers are correct, 6C HZRG's
should be in environments of much lower density than 3C HZRG's. Testing this
is quite difficult. At present the best way of doing this
seems to be to extend the galaxy-counting techniques that have been applied
to the environments of radio galaxies at slightly lower redshifts$^{30,31}$,
but in the near-IR rather than in the optical, since the contrast of any
high-redshift cluster against the unrelated field galaxies is higher
at longer wavelengths. We are currently trying to do this in collaboration
with Hans Hippelein using the OMEGA camera at the Calar Alto Observatory.
In the medium term, the best prospect of testing these answers is 
AXAAF.

If we now return to the first two questions, there seem to me two ways
one can try and answer these. Philip Best and his collaborators$^{25,26}$
suggest that the correlation between radio and near-IR luminosity at
$\rm z \sim 1$ is due to the mass of the central black hole being proportional
to the mass of the galaxy. The bulk kinetic power of radio
jets in 3C galaxies at $\rm z \sim 1$ is close to the Eddington limiting
luminosity of a black hole of mass $\rm \sim 10^8\ M_{\odot}$, and Best
et al. argue that this suggests the bulk kinetic power is primarily
determined by the black hole mass rather than the fueling rate of the
black hole---which is quite plausible given the evidence for  
substantial amounts of gas around HZRG's. In this model, the lack of a 
correlation between radio and near-IR luminosity at lower redshifts
is due to the less plentiful supply of fuel, which means that fuel supply,
rather than black hole mass,
becomes the main determinant of jet power.

Radio luminosity is not just dependent on the bulk kinetic power of the
radio jet, it is also dependent on the density of the surrounding gas$^{32}$,
and I believe it is equally plausible to argue that at $\rm z \sim 1$
radio luminosity depends on stellar luminosity simply because galaxies
that are massive are surrounded by denser, more extended distributions of gas.
I do not have a convincing explanation of why this 
correlation should disappear at low redshift, beyond the speculation that
since the gas around HZRG's is of a very different character to
that around LZRG's (a much larger mass of line-emitting gas at 10$^4$ K),
it is possible that there is a relation between gas mass and stellar mass
at high redshift which disappears at low redshift. 
Finally, I must mention one intriguing point noticed by Nathan
Roche. Although for LZRG's optical luminosity is independent of radio
luminosity over a range of $\rm 10^4$ of radio luminosity, the line
dividing FR2's and FR1's is not independent of optical luminosity$^{20}$. Since
this line has
roughly the same relation as that seen between near-IR and radio luminosity
at high redshift, might this be the fossil of the high-redshift effect? 

\section{Unanswered Questions}

I do not regard any of the questions posed in the last section as being
conclusively answered. I believe the importance of Figure 3 is that
it has suggested questions one stands a chance of answering, questions
which are particularly important because of their connection to
such long-standing fundamental questions about radio galaxies as,
why are low-redshift FR1's found in denser environments than FR2's? What
do radio galaxies evolve into? Why are
low-redshift radio galaxies always giant ellipticals? The 6C and 7C 
samples provide a
basis for obvious observational projects that should go a long way to answering
these questions. First, one would like to put Fig. 3 on a firmer footing.
It has been produced from ground-based imaging, and in the case of the 6C
galaxies, although we are sure that the K-band emission is compact, we can
not put stringent limits on the presence of point sources or even be sure
that the 6C galaxies are ellipticals rather than disk galaxies. With NICMOS
it will be possible to determine how near-IR structure depends on radio
luminosity with much greater certainty. Second, one would like to know
how the relation between near-IR structure/luminosity and radio
luminosity continues to lower radio luminosities. It will be possible
to investigate this by making high-resolution near-IR observations of
the 7C galaxies. Third, one would like to know how other indicators of
the power of the central engine depend on radio luminosity. Two properties
which may (but not definitely) depend on central-engine power are the 
luminosity of the
emission lines$^{14}$ and the strength of the aligned optical component, and
observations of 3C, 6C and 7C galaxies at similar redshifts will show
how both of these depend on radio luminosity (Mark Lacy and 
collaborators have already made a start
on the second of these---this meeting).
Fourth, one would like to know how, at high redshift, radio luminosity depends 
on the density of the environment. The answer here lies in counting galaxies
on near-IR images of 3C, 6C, and 7C galaxies, and, in the longer term, in AXAAF
observations.

Finally, although I have argued that one does not need to justify
one's interest in HZRG's by making a connection to the evolution of
the general galaxy population, I can not resist one speculation.
Figure 3 shows that, at $\rm z \sim 1$, although the rare radio-luminous
3C galaxies are in large elliptical galaxies, most radio galaxies are
in more compact galaxies. Might this not be connected to the apparent
absence of large elliptical galaxies in the Hubble Deep Field$^{33,34}$?

\subsection{Acknowledgements and References}

This research has been a collaborative effort between a large number
of people. I am grateful to (in no particular order)
Gareth Leyshon and Nathan Roche in Cardiff, Katherine Blundell,
Mark Lacy, Steve Rawlings and Chris Willott at Oxford, Mark Neeser at
Groningen, Hans Hippelein at MPIA (Heidelberg), Duncan-Law-Green at Leicester,
Patrick Leahy at Jodrell
Bank, and Garret Cotter at the RGO.

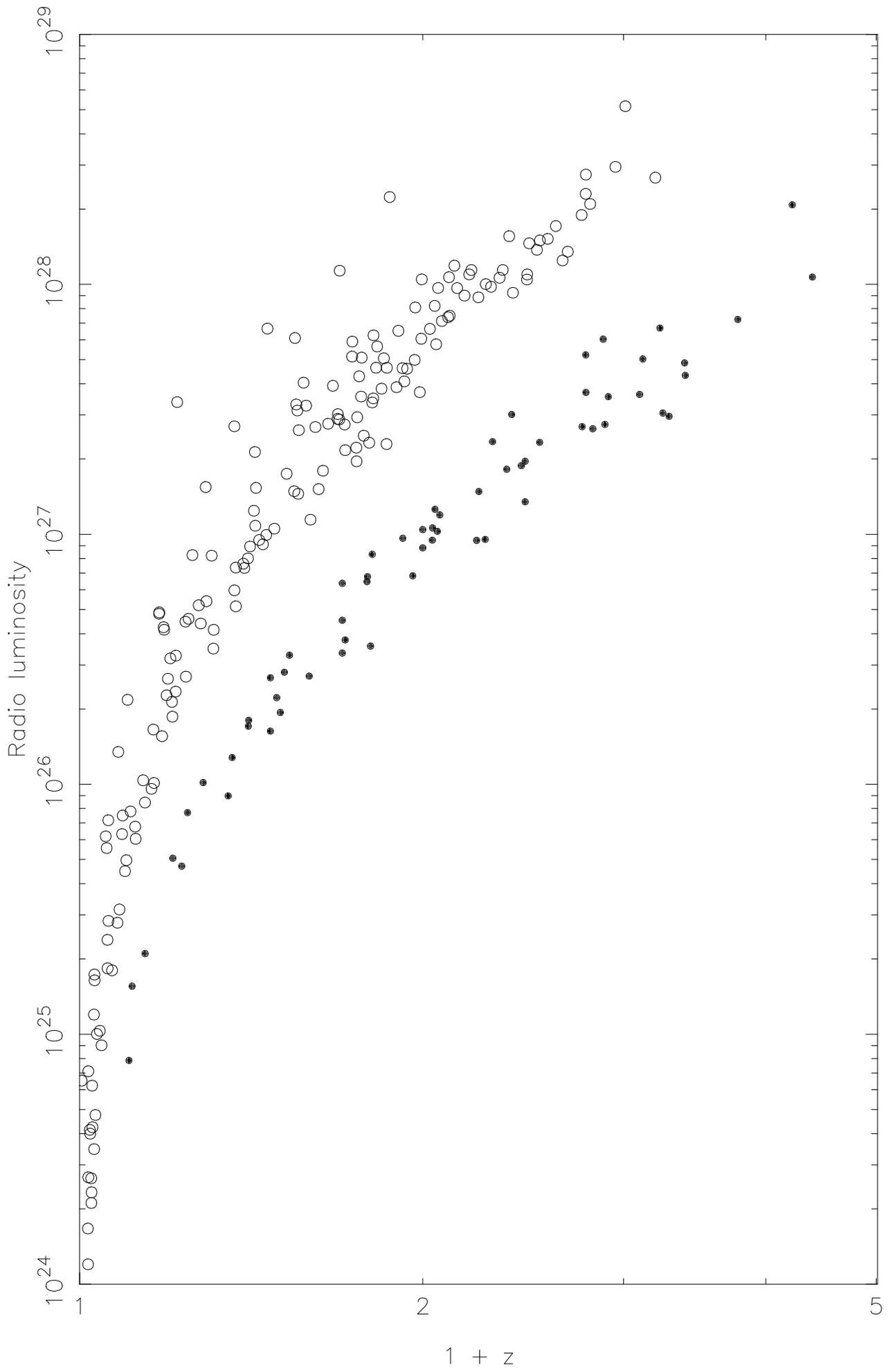
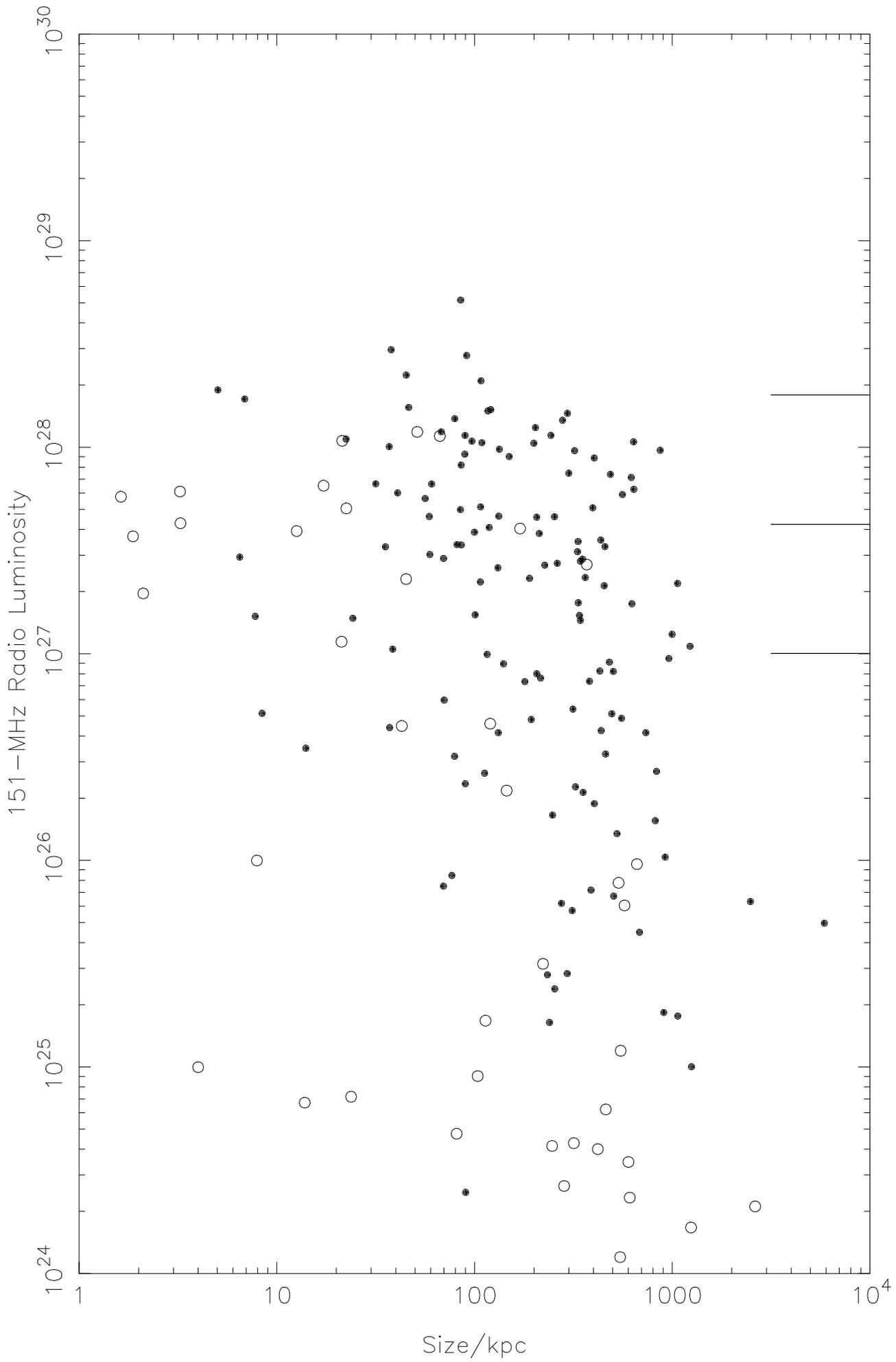
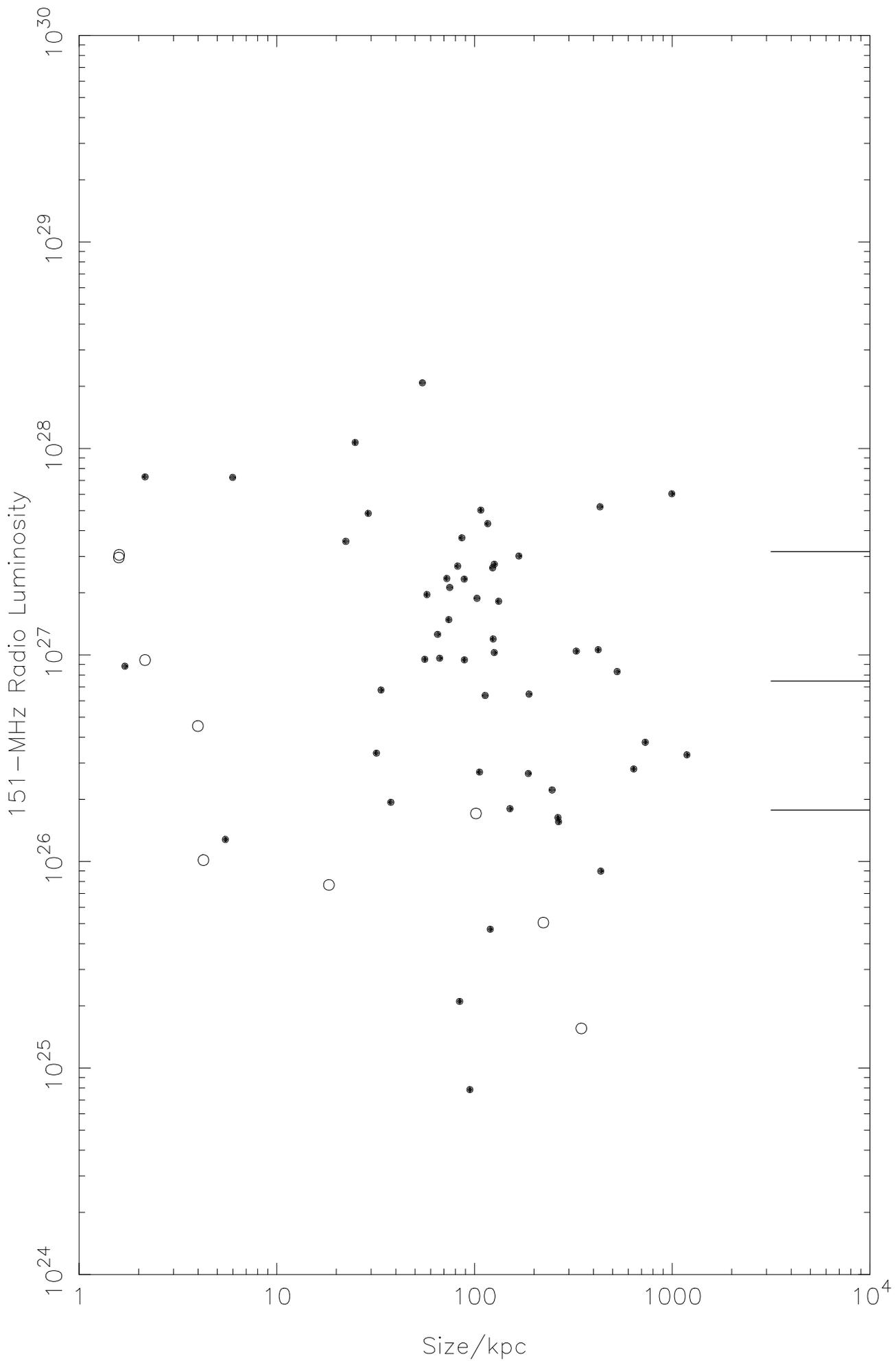
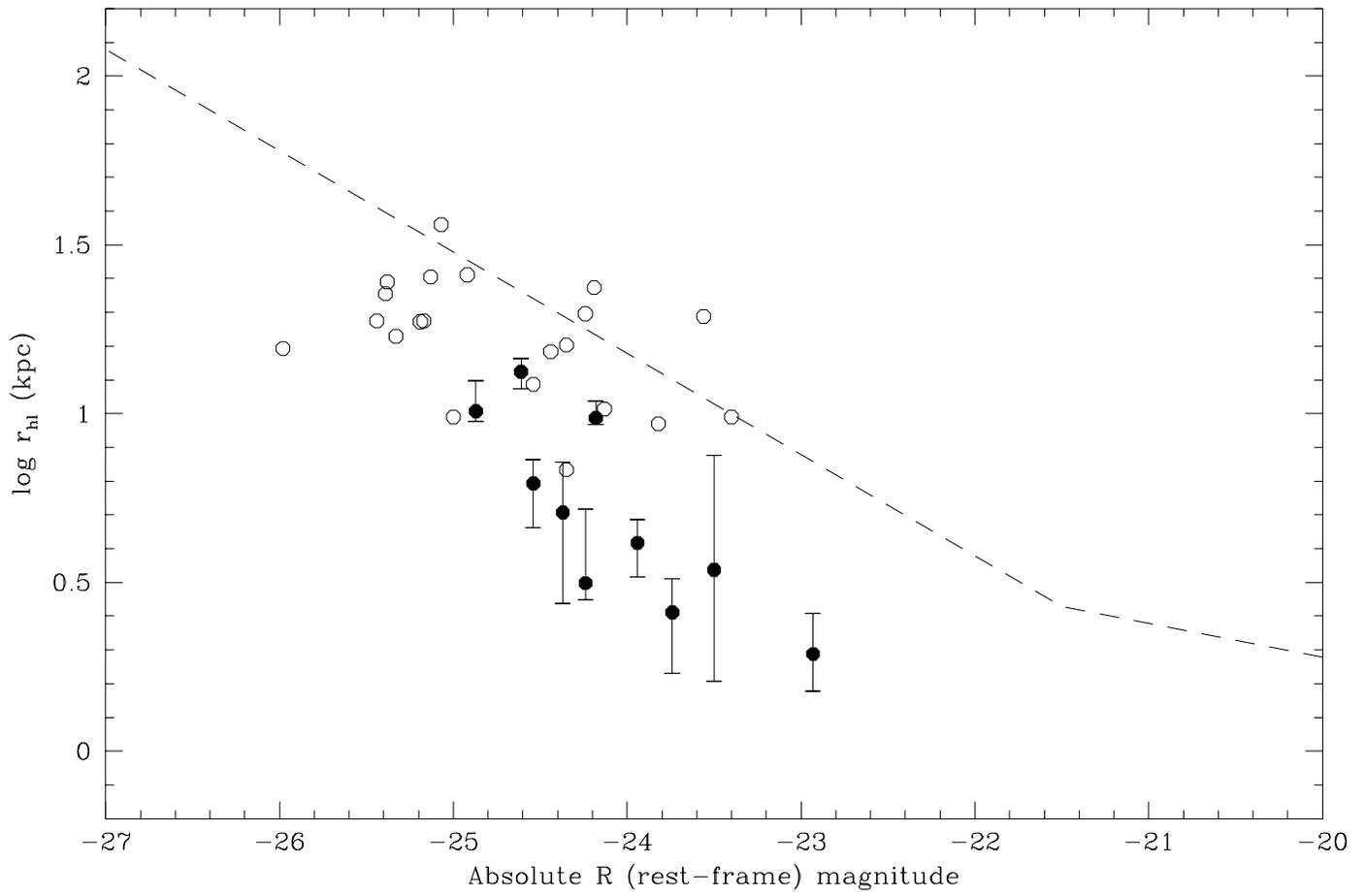
\end{document}